\documentclass[journal]{IEEEtran}
\usepackage{lineno}
\usepackage[T1]{fontenc}
\usepackage{amsmath}
\usepackage{graphicx}
\usepackage{epstopdf}
\usepackage{multicol}
\usepackage{multirow}
\usepackage{xspace}
\usepackage{textcomp}
\usepackage{longtable}

\usepackage{soul}

\usepackage{comment}
\includecomment{hide}

\usepackage{changepage} 

\usepackage{tabularx}

\usepackage{booktabs}
\usepackage{boldline}       
\usepackage{array}       

\usepackage{cellspace}
\usepackage{setspace}
\usepackage[table,xcdraw]{xcolor} 
\usepackage[abs]{overpic} 

\graphicspath{{figs/}}
\newcommand{\fig}[1]{Fig.~\ref{#1}}

\newcommand{\quotes}[1]{``{#1}''}

\begin{document}

\title{Debunking Wireless Sensor Networks Myths}

\begin{hide}
\author{
    Borja~Martinez, 
    Cristina~Cano, 
    Xavier~Vilajosana 
    \thanks{B.\,Martinez and C.\,Cano are with IN3 at Universitat Oberta de Catalunya.}
    \thanks{X.\,Vilajosana is with IN3 at Universitat Oberta de Catalunya and Worldsensing S.L.}%
}
\end{hide}


\maketitle
\IEEEpeerreviewmaketitle

\begin{hide}
\begin{abstract}
In this article we revisit Wireless Sensor Networks from a contemporary perspective, 
after the surge of the Internet of Things. 
First, we analyze the evolution of distributed monitoring applications, 
which we consider inherited from the early idea of collaborative sensor networks. 
Second, we evaluate, within the current 
context of networked objects, 
the level of adoption of low-power multi-hop wireless, 
a technology pivotal to the Wireless Sensor Network paradigm. 
This article assesses the transformation of this technology in its integration into the Internet of Things, 
identifying outdated requirements and providing a critical view on future research directions.

\end{abstract}
\end{hide}

\section{Introduction}
\label{sec:introduction}

Wireless Sensor Networks (WSNs) is a cross-disciplinary research field that has attracted great attention for the last 20 years. 
Initially, WSNs was seen as a fruitful forthcoming research avenue, 
which entailed many challenges in diverse areas such as system-on-chip design, 
low-cost sensors, reduced-size energy storage and low-power networking \cite{Pottie-2000}.

With hindsight, we can question the adequacy of WSNs understood as large deployments monitoring every inch of the environment. Is it reasonably viable to place a device (including a processor, radio transceiver and battery) at each measuring point? Other options such as cameras and satellites offer compelling alternatives for many of the envisioned applications -such as parking and agriculture- in a less-intrusive, environmentally-friendly and sustainable way. 

We also argue that, in the conception of WSNs, the main focus was the technology itself. Questioning whether WSNs were the best solution to the envisioned applications was given far less attention. The research community focused on the technical problems ahead -especially low-power and multi-hop wireless- in order to enable the envisioned large-scale and dense upcoming WSNs. 

However, while the technology was evolving to respond to the initial requirements, real deployments were not materializing as expected. 
The consolidation of the Internet of Things (IoT) paradigm quickly demanded alternative connectivity models. 
Long-range enabled direct Internet access and the collaborative approach of WSNs were no longer central. 
Perhaps for these reasons, 
multi-hop, low-power wireless (which we call \emph{mesh}) did not succeed as expected.
Arguably these emerging long-range models were a response to well-known barriers of \emph{mesh} such as deployment complexity and cost.

Despite this change of priorities, 
the research community is still dedicating substantial efforts on topics 
contextualized in the original WSN framework (large, dense, low-power wireless networks), 
such as routing, medium access, relaying, neighbor discovery and aggregation.
We believe that part of the research community has not realigned their scope. 
This is especially shocking now with the IoT fully consolidated 
but the WSNs approach yet to be seen.
The analysis of the reasons for this misalignment (which may include pressure to publish, inertia, amortization of hardware, lack of self-reflection) are left out of the scope of this article as they require a wider scope into social scientific knowledge.
 


Thus we wonder: are WSNs, understood as large and dense deployments that measure every corner of our surroundings, a myth only believed by WSNs researchers? Have the complex requirements imposed by WSNs hindered the progress of potential applications by unnecessarily complicating the technology?
Based on representative examples, we question the viability of some envisioned WSNs applications. 

Further, we wonder whether focusing on the original WSNs requirements and constraints has made \emph{mesh} networks technologically less competitive against simpler architectures. 
We consider here examples of key technological solutions in the IoT, which offer clear value propositions.
Ultimately, we pose the question: is \emph{mesh} research evolving in the right direction? 

The rest of the paper is organized as follows. 
Section \ref{sec:wsns} overviews the evolution of WSNs, the academic hype and current picture of key applications. 
Then, Section \ref{sec:sensors} delves on the change of requirements of WSN at the application layer.
Section \ref{sec:mesh} analyzes the evolution of \emph{mesh} technology in the context of IoT. We conclude with some final remarks.

\section{The WSN Paradigm and Academic Hype}
\label{sec:wsns}

The WSN paradigm has evolved from the early visionary ideas of what the technology could provide to real-world applications. 
In this section, we argue that academia has been deeply involved in the technology development, but somehow distant from the materialization in the real world.

\subsection{Origins}

In the late 90's, three key technologies made significant advances in reducing size, power and cost: micro-controllers, wireless communications and sensors  
\cite{Kahn-1999}. 
For the first time, compact, autonomous nodes were available for distributed and pervasive outdoor monitoring. 
A new paradigm of 
\quotes{\emph{tiny low-power devices spread over large physical spaces that collaboratively monitor the environment}} \cite{Culler-2004}, the WSNs vision, was taking shape. 

Pister -among other pioneers- drove the fundamentals of WSN research in the years to come under the Smart-Dust project \cite{Kahn-1999}: 

\begin{hide}
    \begin{adjustwidth}{0.6cm}{0.6cm}
    \begin{itshape}
    \vspace*{0.3cm}
    \indent 
    [...] very compact, autonomous and mobile nodes, each containing one or more sensors, computation and communication capabilities, and a power supply. The missing ingredient is the networking and applications layers needed to harness this revolutionary capability into a complete system.
    \end{itshape}
    \end{adjustwidth}
    \vspace*{0.3cm}
\end{hide}

\begin{hide}
    \begin{figure}[ht!]
      \centering
      \includegraphics[width=1.0\columnwidth]{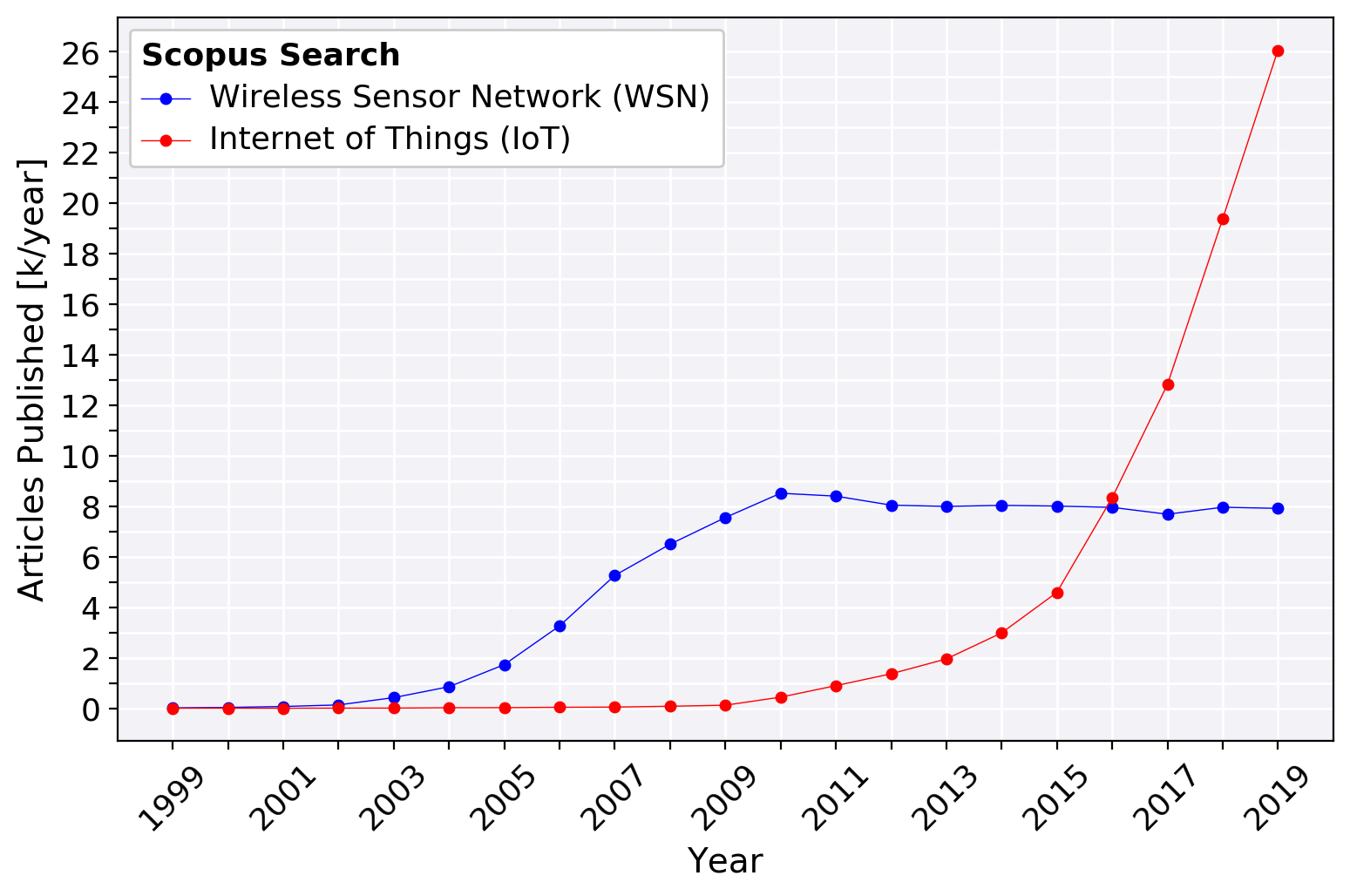}
      \caption{Evolution of publications containing 
      \quotes{WSN} or \quotes{Wireless Sensor Network} in the title, abstract or keywords (source: Scopus, 2020-07-13).}
      \label{fig:f1}
    \end{figure}
\end{hide}

Table \ref{tab:t1} summarizes the vision of the most cited articles in the field. 
We can see that the collaborative idea claimed in \cite{Culler-2004} is common to all. 
However, it is not clear why collaboration was needed: as a way to communicate wirelessly under low-cost/power requirements or for sensing, to benefit from neighboring information. 
In any case, node inter-communication was understood as essential in WSN. 

\subsection{The Academic Hype}

The wireless community felt particularly attracted by WSN research. Significant effort was put into the old challenges that the Internet once had to face (fragmentation, routing, security, etc), but now under hard constraints on low-power operation and computation capabilities.
To get further insight into that hype, note the citations of representative articles on key WSNs communication topics: 
MAC protocols \cite{Ye-2002} (6611 citations), 
architectures and operating systems \cite{Hill-2000} (5122), 
routing \cite{Karaki-2004} (5503), 
security protocols \cite{Perrig-2002} (5537). 
We also show the evolution of WSNs publications (\fig{fig:f1}). 
Note that the WSN hype started around 2003, peaking in 2010 and stabilizing (not decreasing) at around 8k papers/year in 2012. The total has surpassed 100.000 articles in 2019.

Most of the research outputs during this hype focused on technical issues related to networking technologies. However, researchers have been often alien to the increasing gap between the initial expectations and the evolving needs of real-life applications. While the academic proposals were very optimistic in terms of scale and density (see Table \ref{tab:t1}), these were far smaller in real deployments. We review representative examples in the next subsection.

The paradox is that the research community seemed not to worry by this progressive misalignment. The work in \cite{Bhaskaran-2008} is one of the few articles that question the WSNs vision adopted by the academic community, highlighting the lack of depth in the description of application scenarios in research papers. 

Furthermore, we believe the misalignment between WSNs research and real deployments started to be even more evident upon the surge of the IoT (around 2010 in \fig{fig:f1}). 
From then on, the requirements of many applications completely changed under the new paradigm, as we discuss in Sections \ref{sec:sensors} and \ref{sec:mesh}.  

\begin{hide}

\newcommand{\ct}{3.0cm}
\newcommand{\cu}{4.5cm}

\newcommand{\ku}{0.22cm}

\newcommand{\hs}[1]{{\hspace{#1}}}

\begin{table}[ht!]
\centering
\hyphenpenalty=0
\footnotesize
\setstretch{0.9}
\caption{WSN most cited articles (Retrived 2020-04-14)}
\label{tab:t1}
{
\renewcommand{\arraystretch}{1.4}
\begin{tabular}[t]{ p{\ct} @{\hspace{0.2cm}} p{\cu} @{\hspace{0.2cm}} r }
\hlineB{1}
\rowcolor[gray]{.9}
\parbox[c][0.6cm][c]{0ex}{\textbf{Title\,/\,\emph{Highlight}}} &
\textbf{Specifications} &
\textbf{Cites} \\ 
\hlineB{1}
\specialrule{0em}{0em}{0.4em}
\begin{tabular}[t]{@{} p{\ct} @{}}
\raggedright
Akyildiz \textit{et\hs{0.2em}al.}\hs{0.2em}\cite{Akyildiz-2002}\\
\vspace{\ku}
\quotes{\emph{collaborative effort of a large number of nodes}}
\end{tabular}& 
\begin{tabular}[t]{@{}p{\cu}@{}}
\textbf{Scale:} {hundreds-thousands},\\ 
\textbf{Structure:} {dense},\\
\textbf{Lifetime:} {unattended for months-years},\\
\textbf{Size:} {smaller than a cubic centimeter},\\
\textbf{Cost:} {very important.}\\
\end{tabular} &
21696\\ 
\midrule
\begin{tabular}[t]{@{}p{\ct}@{}}
\raggedright
Heinzelman \textit{et\hs{0.2em}al.}\hs{0.2em}\cite{Heinzelman-2000}\\
\vspace{\ku}
\quotes{\emph{local collaboration such as aggregation and fusion}}

\end{tabular} & 
\begin{tabular}[t]{@{}p{\cu}@{}}
\textbf{Scale}: {hundreds-thousands},\\
\textbf{Structure}: {clusterized},\\
\textbf{Lifetime}: {extended},\\
\textbf{Size}: {microsensors},\\
\textbf{Cost}: {relatively inexpensive.}\\
\end{tabular} &
18415 \\ 
\midrule
\begin{tabular}[t]{@{}p{\ct}@{}}
\raggedright
Intanagonwiwat \textit{et\hs{0.2em}al.}\hs{0.2em}\cite{iNTANA-2000}\\
\vspace{\ku}
\quotes{\emph{a collection of nodes coordinate to achieve a larger sensing task.}}
\end{tabular} & 
\begin{tabular}[t]{@{}p{\cu}@{}}
\textbf{Scale}: {hundreds-thousands},\\
\textbf{Structure}: {unstructured sensor fields},\\
\textbf{Lifetime}: {several days},\\
\textbf{Size}: {matchbox sized},\\
\textbf{Cost}: {potential low cost.}\\
\end{tabular} &
8148 \\ 
\midrule
\begin{tabular}[t]{@{}p{\ct}@{}}
\raggedright
Ye \textit{et\hs{0.2em}al.}\hs{0.2em}\cite{Ye-2002}\\
\vspace{\ku}
\quotes{\emph{network of devices collaborate for a common application}}
\end{tabular} & 
\begin{tabular}[t]{@{}p{\cu}@{}}
\textbf{Scale:} {large},\\
\textbf{Structure:} {short-range, multi-hop},\\
\textbf{Lifetime:} {battery-operated},\\
\textbf{Size:} {small},\\
\textbf{Cost:}  {discarded when discharged.}\\
\end{tabular} &
6611 \\ 
\midrule
\begin{tabular}[t]{@{}p{\ct}@{}}
\raggedright
Mainwaring \textit{et\hs{0.2em}al.}\hs{0.2em}\cite{Mainwaring-2002}\\
\vspace{\ku}
\quotes{\emph{nodes cooperate in performing complex tasks}}
\end{tabular} & 
\begin{tabular}[t]{@{}p{\cu}@{}}
\textbf{Scale:} dense,\\
\textbf{Structure:}  multi-hop,\\
\textbf{Lifetime:} 9-12 months,\\
\textbf{size:} {small},\\
\textbf{Cost:} {inexpensive.}\\
\end{tabular} &
5706 \\ 
\midrule
\begin{tabular}[t]{@{}p{\ct}@{}}
\raggedright
Perrig \textit{et\hs{0.2em}al.}\hs{0.2em}\cite{Perrig-2002}\\
\vspace{\ku}
\quotes{\emph{thousands to millions of self-organizing small sensors}}
\vspace{0.13cm}
\end{tabular} & 
\begin{tabular}[t]{@{}p{\cu}@{}}
\textbf{Scale:} {thousands-millions},\\
\textbf{Structure:} {multihop,}\\
\textbf{Lifetime:} {low-power,}\\
\textbf{Size:} {small,}\\
\textbf{Cost:} {inexpensive.}\\
\end{tabular} &
5537 \\ 
\midrule
\begin{tabular}[t]{@{}p{\ct}@{}}
\raggedright
Karaki \textit{et\hs{0.2em}al.}\hs{0.2em}\cite{Karaki-2004}\\
\vspace{\ku}
\quotes{\emph{nodes coordinate among themselves to create a network that performs higher-level tasks}}
\end{tabular} & 
\begin{tabular}[t]{@{}p{\cu}@{}}
\textbf{Scale:} hundreds-thousands\\
\textbf{Structure:} direct/via base station\\
\textbf{Lifetime:} constrained supply,\\
\textbf{Size:} small,\\
\textbf{Cost:} low cost.\\
\end{tabular} &
5503\\ 
\midrule
\begin{tabular}[t]{@{}p{\ct}@{}}
\raggedright
Hill \textit{et\hs{0.2em}al.} \cite{Hill-2000}\\
\vspace{\ku}
\quotes{\emph{networked sensors spread throughout our environment like smart dust.}}
\end{tabular} & 
\begin{tabular}[t]{@{}p{\cu}@{}}
\textbf{Scale:} {numerous},\\
\textbf{Structure:} {multi-hop},\\
\textbf{Lifetime:} {small},\\
\textbf{Size:} {\quotes{one-inch} devices},\\
\textbf{Cost:} {extremely low.}\\
\end{tabular} &
5112 \\ 
\midrule
\end{tabular}%
} %
\end{table}

\end{hide}

\subsection{What Happened to WSNs envisioned applications?}

We revisit here representative WSNs use-cases and discuss their evolution and viability.

\subsubsection*{Unrealistic Applications}

One of the paradigmatic applications of WSNs is habitat monitoring 
(take for instance \cite{Mainwaring-2002} with 5706 citations). 
The initial vision considered thousands of densely deployed sensors that, collaborating with each other, monitor some physical magnitude or phenomenon outdoors. 
In this category fall a good assortment of examples: forest fire detection, water quality monitoring, wildlife tracking, etc. 

In general, these applications are now anecdotal, that is, only relegated to academic deployments and pilots. 
In retrospection, with the current environmental awareness, such WSNs would have involved the deployment of thousands of devices with pollutants that are difficult to recover for recycling. In addition, WSNs may not provide significantly better performance than other systems without the environmental handicap (for example, imaging from satellites). 

However, we believe habitat monitoring is largely responsible for the collaborative vision of early WSNs proposals. Indeed, in some cases, WSNs were even understood as a \emph{single distributed sensor} that monitors a physical magnitude (e.g. temperature of the earth's surface), in contrast to a set of distributed sensors that monitor independent variables 
(e.g. temperature of different machines in a factory). 
In fact, there is a whole research branch based on the physical (co)relations between different measurement points, aiming to exploit compression or prediction. The scarcity of applications of this type that have materialized puts into question the interest of some of these research topics.

\subsubsection*{Non-viable Business Models}

Even when the social, welfare or citizenship benefits are clear, sometimes it seems very difficult to monetize a massive deployment. 
One reason behind the failure to make WSNs profitable has been the complexity and reliability issues associated with deploying and managing large-scale real-world deployments. 
A lesson learnt from academic pilots is the significant human intervention and expertise they required. 
This means that, in practice, the cost of deploying a network was often dramatically higher than the return on investment.

One of the most representative examples under this category is the monitoring of urban outdoor parking. The application seems tailored for multi-hop networks: parking slots evenly spaced by a short distance while data has to travel long distances to the sink. However, the installation of a sensor per spot has not proven a profitable model, at least from the public administrator's perspective. Regulated parking areas tend to maximize occupancy without the need for intervention thanks to the (desperate) private drivers cruising for parking. Considering that the upfront cost of the deployment is afforded by public administrations, and the incremental value for them is marginal with respect to traditional parking policies, no clear ROI exists. This makes the commercialization of such systems difficult. Many companies have been struggling for years to offer viable models for massive deployments of parking solutions. They have not yet succeeded, despite the large number of pilots deployed worldwide.


It may also be that the monitoring of every single parking spot is just disproportionate. The current trend seems to follow a \emph{lean-sensing} approach: monitoring only a few representative spots to provide citizens with simplified/aggregated information (which may be obtained with cameras/radars) about the general state (\quotes{high/low occupation}) of the area. If this model ends up materializing, LPWANs and LTE seem better fitted, since they do not require dense deployments. 



Another use-case common in the literature that falls in this category is smart agriculture (see \cite{Wang-2006}, with 1360 citations). Regardless of the environmental problems, as discussed with habitat monitoring, the payback of a device deployed at each plant seems unfeasible, as the margin for producers is extremely low. We are probably, again, going towards \emph{lean sensing} models, for which long-range networks seem better suited than \emph{mesh}, as we will discuss in Sec.\ref{subsec:lpwan-sensors}. 

\subsubsection*{Alternative Network Solutions}

There are also monitoring applications that have succeeded but using alternatives to \emph{mesh}. In general, applications with low density and/or number of nodes have gradually turned to LPWAN. For example, monitoring a large bridge may require several tens of sensors. Monitoring the landslides of an open pit mine can involve a few hundred. These types of applications -structural health monitoring \cite{Lynch-2006} (1684 citations)- are now mainly using LPWAN. With insight, this may indicate that the original approach was not entirely adequate, especially in terms of scale. This down-scaling may be indicative of an early but significant mistake: promoting density instead of long range transmissions.


\subsubsection*{Success Stories}
Indeed, there are some applications for which the concept of WSN endures, and mesh technologies in particular are actually succeeding. For example, in the chemical, oil and gas factories, most of the flow metering equipment used the Highway Addressable Remote Transducer (HART) protocol. WirelessHART became a natural evolution of that protocol exploiting the concept of wireless mesh networks. Analogously to the smart-parking case, the topology of chemical/oil/gas factories is particularly suitable for \emph{mesh}. The data read from the flow-meters, 
travel over long distances through a physical network of pipes, on which there is some valve or monitoring device every few meters. A key point in this application is that certain levels of reliability are required, and time-synchronized mesh
networks provide the right balance in terms of reliability, density and data-rate. Standardization may also have helped in this case.

Interestingly, this use-case was not envisaged in the early literature. 
Note also that WirelessHART nodes perform a collaborative action to relay information but not at the application layer (e.g., data is not shared, processed nor compressed along the way). 
Another observation we extract from WirelessHART is that successful applications were not created from scratch, wireless (\emph{mesh} in this particular case) just added value over existing solutions. 

It seems that WSN ideas found their opportunity in well-established scenarios, but not in the most paradigmatic cases envisioned in the early literature.

\begin{hide}
  \begin{figure}[ht!]
    \centering
    \includegraphics[width=1.0\columnwidth]{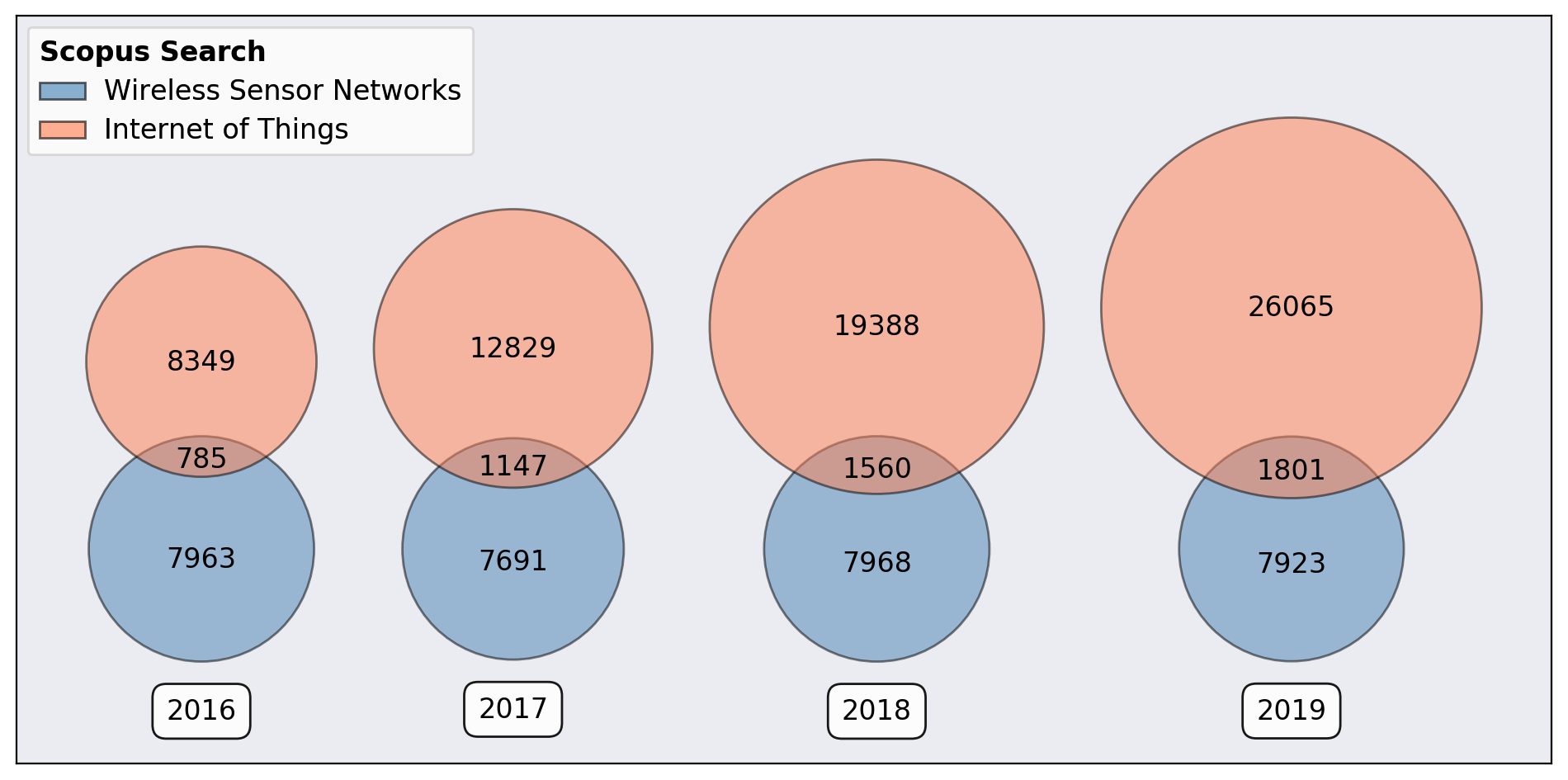}
    \caption{Number of articles published between 2016 and 2019 that include 
    \quotes{Wireless Sensor Networks} in the keywords (bottom), \quotes{Internet-of-Things} (top) and those that include both terms (intersection). The latter can be indicative of authors that consider WSN now as a part of the IoT.}
    \label{fig:f2}
  \end{figure}
\end{hide}

\section{From sensor networks to networked sensors}
\label{sec:sensors}

\begin{hide}
  \begin{figure*}[ht!]
    \centering
    \includegraphics[width=0.9\textwidth]{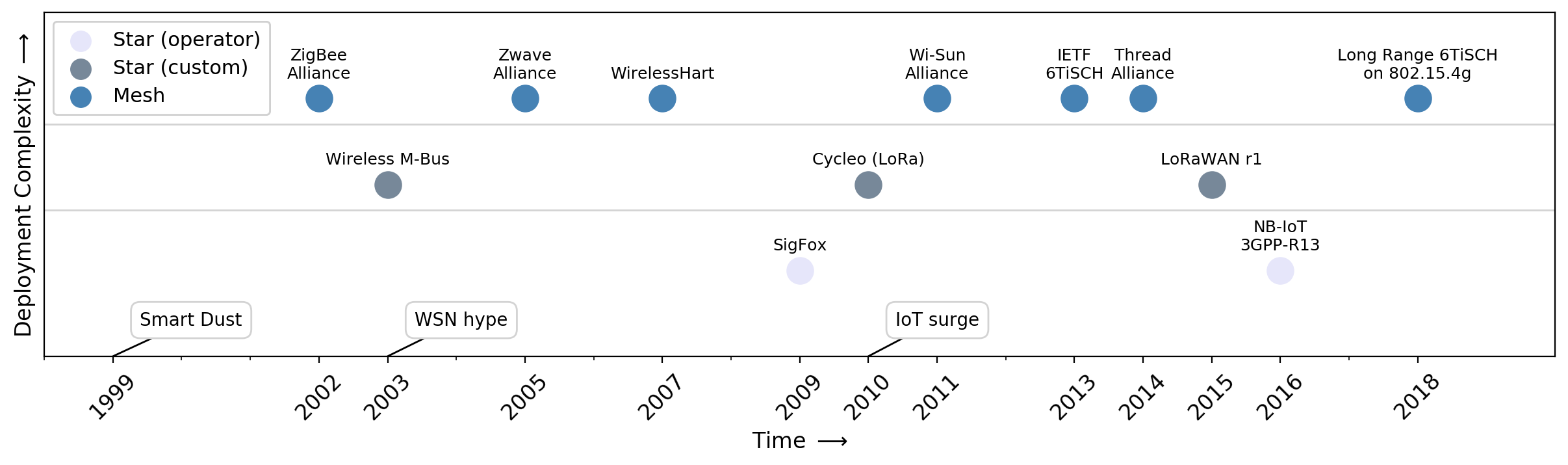}
    \caption{A timeline of considered wireless technologies and their complexity of deployment from adopters perspective.}
    \label{fig:f3}
  \end{figure*}
\end{hide}

Today, some researchers consider WSNs somehow diluted in a broader paradigm, 
the IoT (see \fig{fig:f2}). However, the IoT entails fundamental changes from the WSN initial vision. 
In most IoT applications inter-communication between end devices is not strictly required at the application level. 
In contrast, for an IoT device, the only essential requirement is an Internet connection. 
And if some kind of collaboration between end devices is required, in the current paradigm it will most likely be done at the Cloud.

However, it is this dependence on neighboring nodes to relay information 
-which requires to have at least one node in sight- 
that has generated important issues in \emph{mesh} deployments. 
This is also the reason why applications in which sensors are scattered (low-density) or even isolated (stand-alone) are particularly unsuited for \emph{mesh}. Examples of this, widely referred in the literature, are infrastructure (e.g. a dam), environmental (e.g. water quality of a river) and seismic (e.g. volcano's activity) monitoring. 

In this section, we discuss the transformation of WSN monitoring applications and we delve into the role of remotely-readable sensors in the IoT context.

\subsection{Low-density networks: the rise of LPWAN}
\label{subsec:lpwan-sensors}

The expectations of WSNs lead to network models characterized by short-range radios, which may require placing relays 
at specific positions to ensure connectivity in the field. High density expectations lead to a model in which high density was a necessity, and ensuring density impacts the complexity and cost of deployments. This opened a perfect opportunity for long-range single-hop technologies and many applications pivoted to them.

Nowadays, for non-dense deployments, 
LPWAN are preferred (in 2019, Semtech reported more than 100M LoRa devices), even when considering the bandwidth and duty cycle limitations. We argue that the main reason is the lower deployment complexity provided by long-range technologies. Indeed, in LPWAN there is no need to worry about the specific position of each device as data is directly sent to the back-end. 

Looking ahead, LPWAN is now converging with \emph{mesh}. Long-range mesh may be a way of being more competitive in some markets (we will see the case of utility metering in the next section). However, it can be of little use for applications with deployments isolated from each other (e.g. infrastructure monitoring).


We also believe that developing the concept of \emph{mesh} on LPWAN may make itself overshadowed by simpler alternatives, 
such as Narrowband Internet of Things (NB-IoT). 

In \fig{fig:f3} we show the timeline of the technologies described
in this article according to their deployment complexity.
Observe the periodic trend of complex approaches followed
by simpler ones that disrupt the market.


\subsection{Stand-Alone Sensors and Devices: a scenario for cellular}
\label{subsec:lte-sensors}
 
The IoT is mostly composed by stand-alone devices, for which cellular solutions are far more appropriate. The examples are countless. Vending machines incorporate LTE cards for monitoring stock. Vehicles are factory-equipped with LTE to offer monitoring services for the driver (battery, tire pressure, door lock...) and the manufacturer (predictive maintenance). 
Urban shared electric bikes/motorbikes use LTE to report their geolocation. 

For these applications,
forming an interconnected network is difficult. 
\quotes{Things} exist independently of each other, sometimes isolated permanently (e.g. a weather station) or intermittently (e.g. logistics).  The formation of sparse networks is an open issue for \emph{mesh} research, in our opinion more urgent than focusing on ultra-dense network management.

Mobile objects deserve special attention. For objects that follow a predefined path (e.g. urban public transport), \emph{mesh} may be technically viable but, at the moment, only globally operated networks can provide ubiquitous coverage. 
For objects that require coverage in positions that cannot be determined in advance (e.g. private vehicles), global networks are probably the only option. 
A representative example is
cattle tracking, with multiple companies commercializing LTE-based solutions.

\section{The evolution of mesh within the IoT }
\label{sec:mesh}

In this section, we discuss the technological evolution of \emph{mesh}, 
a technology closely related to the WSN paradigm that is now evolving in the IoT context. 

Even though the concept of \emph{mesh} dates back to the late 1990s, the first \emph{mesh} standards  appeared around 2003 (see \fig{fig:f3}),
partly as a response to the demands of WSN pioneers.
As we have mentioned, soon the requirements envisioned by the research community and industries started to diverge, and the gap increased with the rise of the IoT.  
While \emph{mesh} research was entangled in increasingly complex issues under the original large-scale premise, 
some IoT use-cases were 
moving in a different direction. 
In the new paradigm we identify two main trends. First, a progressive de-escalation, a downgrade of the large-scale premise of WSNs. Second, fragmentation, a clustered and hybrid landscape -- as opposed to the flat, non-hierarchical approach of the early settings. 
In the IoT arena, far from the original specifications, \emph{mesh} found a tough competition with other wireless technologies. 

\subsection{Micro networks}
\label{subsec:micro-mesh}

Smart-home and, more generally, building automation has been a primary target for \emph{mesh} from the early days of Zigbee. 
While serious doubts about the ZigBee performance started to arise, Z-Wave (Zensys) emerged as a simpler -and less expensive- alternative. One of the noteworthy aspects, regarding our discussion, is the downscaling in the number of addressable devices, from 65536 in ZigBee to 252 in Z-Wave. At the beginning of 2019, the Z-wave alliance announced a market share of 100M+ devices, with more than 2600 certified products, a number that has significantly grown in recent years. This good market momentum seems to indicate that the limit of 252 devices is not perceived as a handicap.

Mesh reemerged in smart-home with the launch of Thread in 2014, a mesh protocol promoted by Samsung, ARM and Nest (Google), among others. Thread uses 6LoWPAN on top of IEEE 802.15.4 and mesh routing is implemented through a flavour of the Routing Information Protocol (RIP). There are two aspects of Thread revealing of current trends. 

First, as in the case of Z-Wave, the maximum number of end-devices is limited to 511 per router. Networks support up to 32 routers, but it seems unrealistic to have many of them in a domestic environment
(see \fig{fig:f4}). 
Furthermore, sleepy end-devices (those battery powered) 
communicate only through their parent router, not between them.
This is radically different to fully-meshed WSN inter-device communication approach.

Second, Thread is not seen as a replacement for WiFi. 
Rather, it is envisaged as a coexisting complement, providing IPv6 to products or use-cases not suitable for WiFi
(due to power and/or cost reasons).
Home devices become virtualized objects, decoupled from the connectivity technology used.
Note the IoT flavour: to control some device (e.g. a thermostat) from a smart-phone application, both smart-phone and device are connected to the Internet (most likely to a Cloud service) where the operation is actually executed. This is also different to older approaches such as those based on Bluetooth, in which the device connects directly to the smart-phone via pairing. 

Emerging mesh technologies such as 6TiSCH, which try to position in the building automation space, have been designed to operate in dense networks 
under strict low-power requirements. 
These foundational requirements seem not aligned to the reality of current deployments.

To sum up, from a research perspective, in indoor and domestic scenarios we observe a shift. 
Mesh networks 
are scaled down to small clusters that forward data to internet-connected hubs. 
In this scenario, part of the research developed within the WSNs context (e.g. \emph{compressive sensing}, \emph{over-the-air compression}, \emph{asynchronous sleep-based MAC protocols}...)  may not be applicable, as it is not suited to small, cloud-centric indoor deployments.

\begin{hide}
  \begin{figure}[h]
    \centering
    \includegraphics[width=1.0\columnwidth]{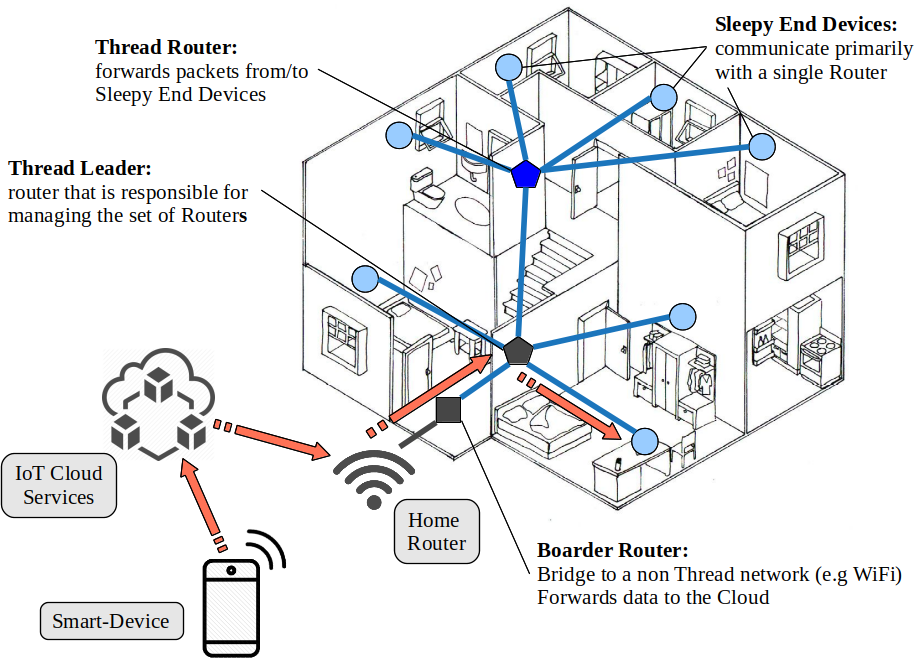}
    \caption{Typical smart-home setting}
    \label{fig:f4}
  \end{figure}
\end{hide}

\subsection{Technological Fragmentation}
\label{subsec:fragmented-mesh}
In smart-home scenarios we have stressed the coexistence of several technologies forming small clusters of devices confined in a reduced space, which may eventually form a micro-mesh network. The main players (Google, Samsung...) are fully aware of this fragmentation, as evidenced by the multi-standard smart-hubs they currently offer in the market. 
These routers support traditional star topologies such as WiFi and BLE, but can also handle mesh protocols such as Zigbee, Z-Wave and Thread. 
There is no clear dominant standard, different technologies just coexist. 

A different case is the Automatic Meter Reading (AMR). Remote access to utility meters delivered the ability to reduce the intervals between readings from months to minutes, thus enabling new services for users and especially for operators, particularly interested in balancing loads. The value proposition is clear and the market huge, both potential and actual.

The complexity of the AMR market 
relies on the fact that it can be addressed with multiple technologies, 
all of them having advantages and drawbacks. On the one hand, the number of utility meters in a building may not justify a mesh network. 
On the other hand, it is perhaps too expensive to connect each meter using LTE. But the typical arrangement of the utility meters in a building makes it straightforward to reach a shared access point and, from there, connect to the backbone using LTE, resulting in hybrid solutions.

It should be emphasized that AMR services are intrinsically individual: the remote-operated meters do not need communication between them. This makes AMR also compatible with the LPWAN single-hop approach, although ensuring the coverage for all customers is still an open issue. The latter may be feasible in urban settings, but in rural areas bridges to other networks will often be necessary. Finally, 3GPP has their own proposal, NB-IoT, which seems particularly suitable for AMR. As NB-IoT shares the cellular network, it offers a good solution to the aforementioned problem of LPWAN coverage. In the case of energy meters, power-line communication (PLC) can be used to transmit the readings.

All these options available have lead to a very fragmented market. For energy, many of the meters are remotely-operated through PLC, mainly using the PRIME standard, an alliance that announces more than 20 million remote-meters deployed worldwide. Obviously, PLC is particularly suitable for this type of meters. 
For water and gas the market is more complex. 
Europe has adopted the Wireless M-Bus standard (13757-x).  
Wireless M-Bus is based on a star topology
operating at 868MHz in T or C mode, and 168MHz in N mode 
for difficult radio environments.
Still, proprietary solutions are widely used.


Another prominent actor in the AMR market is the WiSUN alliance. 
It is focused on providing mesh connectivity for utility meters 
Major vendors joined the alliance and has a strong level of adoption in Asia. WiSUN reports +95M devices running worldwide.

Meanwhile, manufacturers remain expectant about the course of events with LPWAN. 
Most manufacturers offer versions of their meters compatible with SigFox, LoRa and similar technologies. Up to today there is no significant LPWAN momentum in this market, but the technology is mature, ready for immediate use.
In the midst of this competitive landscape, mobile operators are already deploying NB-IoT, although the general feeling is that NB-IoT arrives too late.

In summary, from a research perspective, \emph{mesh} is well positioned in AMR. 
However, it is noteworthy that AMR is far from original WSN proposals. In fact, the IoT flavour is remarkable again: the sole objective is to push small pieces of data (counts) into the Internet. There is no need for interconnection between meters, no opportunity for collaboration; the possibility of compression is scarce or non-existent. As in the smart-home use-case, much of the research carried out within the WSNs framework can hardly be fit in this IoT context.

Still, there is an open question regarding \emph{mesh} and AMR. 
As we have seen, recent trends are evolving towards long-range \emph{mesh}. 
In this use-case, this could indeed enable joining together several buildings in the same (macro) mesh, 
forming a network that is then connected to the Internet. 
Alternatives such as connecting each building to the Internet may be simpler and more cost-effective. 
The evolution of WiSUN is yet to be seen.


\section{Final Remarks}
\label{sec:conclusion}

In this article we revisit, from the current IoT perspective, the evolution of WSNs, a research paradigm proposed 20 years ago that has mobilized vast research efforts.

The article first reviews WSNs use cases with a critical thinking perspective, analyzing what is their level of maturity and adoption today. In particular we identify some applications that never materialized, although they influenced the specifications of WSNs-enabling technologies. 

In this article we also study the impact of the technologies developed within the WSN conceptual framework. First, the evolution of distributed monitoring applications (networks of sensors) is analyzed. In particular, we see that the concept of collaboration between sensors has rarely been used in current monitoring applications. 
In contrast, nowadays sensor networks tend to be sparse, with sensors directly communicating to a gateway.
Often sensors are completely isolated. 
We claim that sensor networks have evolved to  networked sensors. 
We have shown that this may explain why these applications have pivoted to LPWAN or cellular technologies.

Finally, we assess the adoption of low-power wireless multi-hop networks (\emph{mesh}), 
a radio technology used today in some IoT verticals. 
We note that the initial requirements that lead to the development of mesh networks diverge from the needs of some popular IoT applications. In particular, we observe a scale and density reduction leading to clustered micro-network structures. We also point out that mesh networks are used in verticals such as industrial automation, smart homes and AMR, but coexisting with other technologies in very fragmented markets. 

The motivation of this article is to reassess the considerations on which future research should be based. We observe still today numerous research efforts ($\sim$8k papers/year) contextualized with requirements inherited from the pioneering literature of WSNs, which may no longer apply. This indicates a potential misalignment between academic research and applicability.

As a final remark, WSN research has launched or stimulated many cross-disciplinary topics. 
The decline of the original paradigm does not necessarily imply the loss of interest in some of these topics. 
Among them, it is worth noting localization and time-synchronization, still open-issues of full interest.




\begin{hide}
  \bibliographystyle{IEEEtran}
  \bibliography{commag-paper}
\end{hide}

\vfill
\newpage

\begin{IEEEbiographynophoto}{Borja Martinez}
received his B.Sc. in physics 
and the Ph.D. in informatics from the Universidad Aut\'onoma de Barcelona (UAB), Spain, 
where he was assistant professor from 2005 to 2015. 
He is currently a research fellow at the IN3-UOC. 
His research interests include low-power wireless technologies and energy management policies. 
\end{IEEEbiographynophoto}

\begin{IEEEbiographynophoto}{Cristina Cano}
holds a Ph.D. degree in communications from Universitat Pompeu Fabra in 2011. 
She has been a research fellow at the Hamilton Institute of Ireland, Trinity College Dublin and Inria. 
She is now an associate professor at UOC. 
Her research interests include the coexistence of wireless networks, distributed resource allocation, and online optimization. 
\end{IEEEbiographynophoto}

\begin{IEEEbiographynophoto}{Xavier Vilajosana}
received his B.Sc. and M.Sc in Computer Science from Universitat Polit\`ecnica de Catalunya 
and his Ph.D. in Computer Science from the Universitat Oberta de Catalunya. 
He has been a researcher at Orange Labs, at UC Berkeley. 
He is now professor at UOC.
His research interests include wireless communications and standardization.
\end{IEEEbiographynophoto}


\vfill

\end{document}